\documentstyle[11pt]{article}
\begin{document}
\mbox{\hspace{6cm}To be published in {\it Comments on Condensed Matter Physics}}
\begin{center}
{\LARGE{\bf Bridging the length and time scales: from {\em ab initio}
electronic structure calculations to macroscopic proportions}}
\vskip 1.5em
{\large \lineskip .5em \begin{tabular}[t]{c}
Paolo Ruggerone, Alex Kley, and Matthias Scheffler\\
Fritz-Haber-Institut der Max-Planck-Gesellschaft,\\
Faradayweg 4-6, D-14\,195 Berlin-Dahlem, Germany\\
\end{tabular}\par}
\end{center}
\begin{abstract}
Density functional theory (DFT) primarily provides a good description
of the electronic structure. Thus, DFT {\em primarily} deals with
length scales as those of a chemical bond, i.e. 10$^{-10}$~meter, and
with time scales of the order of atomic vibrations, i.e. 10$^{-13}$
seconds.  However, several interesting phenomena happen
and/or become observable on different scales, namely meso- or macroscopic
lengths and on time scales of seconds or even minutes.  To bridge the
gap between 10$^{-13}$~seconds and a second or between
10$^{-10}$~meter and 10 and more nano meters is one of the important
challenges we are facing today.  In this paper we show how we are
overcoming these time and size problems for the example of crystal growth
and the evolution of nano-scale structures. The key is a kinetic Monte
Carlo approach with detailed input from DFT calculations of the relevant
atomistic processes.
\end{abstract}

\section{Introduction}\label{intro}
Many, maybe most, interesting physical phenomena take place with meso- or
macroscopic length scales and over times that range from seconds to minutes. For
example, surface reconstructions sometimes evolve over a time period of seconds
or even minutes, and the self-organization of nano-scale structures, such as
for example quantum dots, also occurs over macroscopic times. {\em Ab initio}
calculations (electronic structure, total energy calculations as well as
molecular dynamics (MD) simulations) are concerned with length scales of a
chemical bond and with times determined by interatomic force constants and the
corresponding atomic vibrations. To bridge the gap {\it from the atomistic
  processes to macroscopic dimensions} is an important challenge.  In this
paper we show how we are coping with this issue for the example of crystal
growth by means of a kinetic Monte Carlo (KMC) approach based on detailed
input from density functional theory (DFT) calculations. This approach is able
to describe the evolution of meso- and macroscopic growth shapes which
may (see below) differ significantly from equilibrium shapes as determined by
the minimum of the free energy.

The growth conditions, namely substrate temperature and deposition
rate, and the presence or absence of surface defects and impurities
play the relevant roles for crystal growth far from thermal
equilibrium.  For example, according to STM studies Pt islands on
Pt\,(111) show fractal-like structures, triangular
and hexagonal shapes at different substrate temperatures~\cite{michely93}. Moreover,
different densities of Pt islands coexist on reconstructed and
unreconstructed terraces on Pt\,(111)~\cite{hohage95}.  Furthermore, a
small increase in the temperature produces a transition in the
shape of Ag islands on Pt\,(111) from fractal to
dendritic~\cite{brune94}.  It is clear that a thorough knowledge of the atomistic
processes occurring at surfaces during growth is a fundamental
requirement for understanding these phenomena.

Our strategy for gaining insights into the interplay of microscopic processes
in assembling mesoscopic structures is based on the combination of DFT
calculations and KMC simulations. This approach is called the ``{\em ab initio}
kinetic Monte Carlo method'' and proceeds in three steps:
\begin{itemize}
\item[1)] Analysis of all possibly relevant atomistic processes using
DFT.
\item[2)] Selection of the relevant atomistic processes and evaluation
of their energy barriers and prefactors.
\item[3)] A KMC study using the parameters determined in the second step.
\end{itemize}
The main risk of this approach lies in step 1), namely that an
important process is overlooked. However, as step 1) may be combined
with an {\em ab initio} molecular dynamics study, this risk is rather
low.  Altogether the {\it ab initio} KMC approach is able to describe
the evolution of structures of 10, or 100, or more nano meters in
dimensions, and to cover time scales of seconds. This is done without
introducing any additional significant approximation than that of the
exchange-correlation functional contained in the DFT step. 
Technical approximations, as for example supercell size and/or {\bf
k}-point summation, will be relevant as well, but these could (and
should) be checked by performing the necessary tests.

The paper is organized as follows. In the next Section we will give a brief
picture of processes involving an adatom that has landed on a
substrate. In Section~\ref{kmc} we describe the main features of the ``{\em ab
  initio} kinetic Monte Carlo method'' (more details are contained in
Ref.~\cite{ratsch97}). Section~\ref{risultati} contains our results for Al on
Al\,(111) and, particularly, a discussion of the shapes of islands and their 
evolution with time.

\section{Atomistic Processes}\label{atompro}
\begin{figure}[t]
\unitlength1cm
 \begin{center}
    \begin{picture}(10,6.5)
      \includegraphics{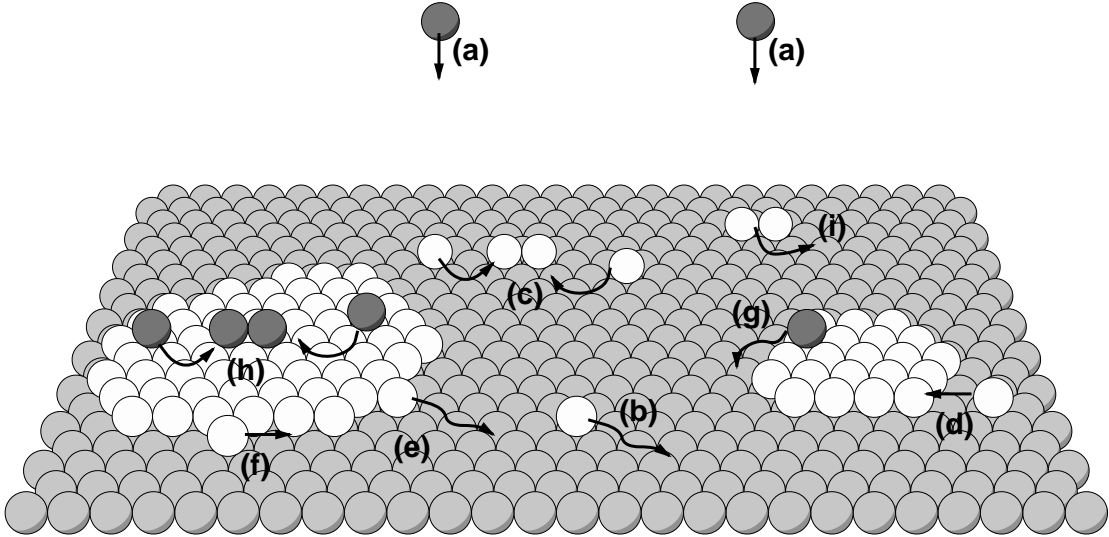}
    \end{picture}
 \end{center}
\caption{
The different atomistic processes for adatoms on a surface:
$(a)$ deposition,
$(b)$ diffusion at flat regions,
$(c)$ nucleation of an island,
$(d)$ diffusion towards and capture by a step edge,
$(e)$ detachment from an island, 
$(f)$ diffusion parallel to a step edge,
$(g)$ diffusion down from an upper to a lower terrace,
$(h)$ nucleation of an island on top of an already existing island, and
$(i)$ diffusion of a dimer (or a bigger island).
For the processes $(a), (c), (g)$ and $(h)$ also the reverse
direction is possible, but typically less likely.}
\label{processes}
\end{figure}

During growth an atom that reaches the surface, may either stay on the
surface and wander around, form an island with other wanderers, or
evaporate back into the gas phase. Under typical growth conditions the
latter is a very rare process and can be neglected.  Thus, we focus
our attention on the wandering adatom that can be involved in several
processes. These are schematically illustrated in
Fig.~\ref{processes}.  After deposition $(a)$ atoms can diffuse across
the surface $(b)$ and will eventually form a small nucleus when they
meet another adatom $(c)$, or adatoms get captured by an already existing
island or a step edge $(d)$. Once an adatom has been captured by an
island, it may either leave the island and return to the flat surface
({\it reversible aggregation}) $(e)$ or remain bonded to the island
({\it irreversible aggregation}). An atom that is bonded to an island
may diffuse along its edge until it finds a favorable site $(f)$.  At
low coverage of adsorbed material (say $\Theta \le 10~\%$), deposition
on top of islands is insignificant and nucleation of islands on top of
existing islands practically does not take place. However, if the step
down motion $(g)$ is hindered by an additional energy barrier (often
called Schwoebel-Ehrlich barrier~\cite{stepedge}), nucleation of
island on top of islands becomes likely $(h)$, giving rise to rough
surfaces.  A negligible step-edge barrier favors an easy motion of an
adatom from the upper to the lower terrace with a consequent growth of
smooth films.
In principle it is possible that not just single adatoms but
also dimers and bigger islands migrate $(i)$. 
For example, a dimer might diffuse by the two
atoms rotating around each other. Moreover, compared to a single adatom, a
dimer may be less bounded to the substrate since the electrons of the two
adatoms participate to the adatom-adatom bond and not only to the
adatom-substrate bonds. Therefore, it may be expected a low activation
barrier for the diffusion of dimers, but there is no clear evidence yet available. 

In the case of surface diffusion we have assumed that an adatom hops from one
binding site to another ({\it hopping}), but this is not the only way: An
adatom may diffuse by atomic exchange where it changes place with a substrate
atom and the ejected substrate atom moves further ({\it exchange}).  This
mechanism (first discussed by Bassett and Webber~\cite{bas78} and Wrigley and
Ehrlich~\cite{wri78}) is actuated by the desire of the system to keep the
number of cut bonds low along the diffusion pathway.  On fcc\,(100) surfaces
diffusion by atomic exchange was observed and analyzed for Pt~\cite{kel90} and
Ir~\cite{che90}, and predicted for Al~\cite{feibelmann90}, Au, and strained
Ag~\cite{yu97}.

\begin{figure}[t]
  \leavevmode \includegraphics{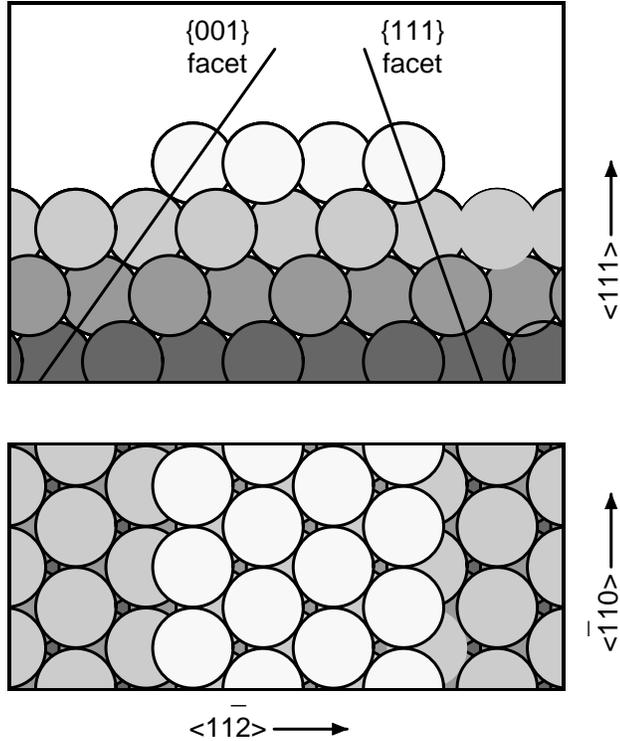} 
\vspace*{10cm}
\caption{
Top view of the two types of close-packed steps on the \,(111) surface
of an fcc crystal. The left step has a $\{100\}$ facet (square geometry) and
the right one has a $\{111\}$ facet (triangular geometry).}
\label{ste111}
\end{figure}

Diffusion along a step edge may also occur via the exchange mechanism
as on the Al\,(111) surface. Since this system will be the object of our
study, we add here a few more words. As illustrated in
Fig.~\ref{ste111} on the \,(111) surface of an fcc crystal there are
two different types of close-packed steps.
They are labeled according to their $\{100\}$ and $\{111\}$ facets,
referring to the plane passing through the adatom of the step and the
neighbor atom of the substrate (often these two steps are labeled A
and B, respectively).  On Al\,(111) DFT calculations~\cite{stumpf94}
show that an adatom at the $\{111\}$ faceted step experiences a rather
high diffusion barrier, if the mechanism would be hopping: Either it
has to move over an ontop site of the substrate or to leave the step edge to
finally reach an adjacent step edge position.  Because this is 
energetically unfavorable, the system chooses an alternative: an 
{\it in-step} atom moves out of the step and the adatom fills the opened site. 
Thus, the coordination of all the particles does
not decrease appreciably during the whole process, and the
corresponding energy barrier is lower than that of the hopping
process. Yet, the final geometry is indistinguishable from that of a 
simple hop. Along the $\{100\}$ facet the Al adatom prefers to jump since
the associated path goes through a bridge position with a consequent
lower activation energy than for an exchange. We will see that this
difference will play an important role for understanding the shapes of
the islands during growth.

\begin{figure}[t]
  \begin{center}
    \unitlength=1.cm
    \begin{picture}(12,5)
      \thicklines
      \begin{picture}(2,0)
        \put(.400,3.200){\circle{.5}}
        \put(.900,3.200){\circle{.5}}
        \put(1.400,3.200){\circle{.5}}
        \put(1.900,3.200){\circle{.5}}
        \put(2.400,3.200){\circle{.5}}
        \put(2.900,3.200){\circle{.5}}
        \put(.650,3.633){\circle{.5}}
        \put(1.150,3.633){\circle{.5}}
        \put(1.650,3.633){\circle{.5}}
        \put(2.150,3.633){\circle{.5}}
        \put(2.650,3.633){\circle{.5}}
        \put(3.150,3.633){\circle{.5}}
        \put(.400,4.066){\circle{.5}}
        \put(.900,4.066){\circle{.5}}
        \put(1.400,4.066){\circle{.5}}
        \put(1.900,4.066){\circle{.5}}
        \put(1.650,4.499){\circle*{.5}}

        \put(5.400,3.200){\circle{.5}}
        \put(5.900,3.200){\circle{.5}}
        \put(6.400,3.200){\circle{.5}}
        \put(6.900,3.200){\circle{.5}}
        \put(7.400,3.200){\circle{.5}}
        \put(7.900,3.200){\circle{.5}}
        \put(5.650,3.633){\circle{.5}}
        \put(6.150,3.633){\circle{.5}}
        \put(6.650,3.633){\circle{.5}}
        \put(7.150,3.633){\circle{.5}}
        \put(7.650,3.633){\circle{.5}}
        \put(8.150,3.633){\circle{.5}}
        \put(5.400,4.066){\circle{.5}}
        \put(5.900,4.066){\circle{.5}}
        \put(6.400,4.066){\circle{.5}}
        \put(6.900,4.066){\circle{.5}}
        \put(6.900,4.566){\circle*{.5}}
        \put(6.900,4.566){\vector(1,0){.55}}

        \put(10.400,3.200){\circle{.5}}
        \put(10.900,3.200){\circle{.5}}
        \put(11.400,3.200){\circle{.5}}
        \put(11.900,3.200){\circle{.5}}
        \put(12.400,3.200){\circle{.5}}
        \put(12.900,3.200){\circle{.5}}
        \put(10.650,3.633){\circle{.5}}
        \put(11.150,3.633){\circle{.5}}
        \put(11.650,3.633){\circle{.5}}
        \put(12.150,3.633){\circle{.5}}
        \put(12.650,3.633){\circle{.5}}
        \put(13.150,3.633){\circle{.5}}
        \put(10.400,4.066){\circle{.5}}
        \put(10.900,4.066){\circle{.5}}
        \put(11.400,4.066){\circle{.5}}
        \put(11.900,4.066){\circle{.5}}
        \put(12.400,4.086){\circle*{.5}}

        \put(.400,.200){\circle{.5}}
        \put(.900,.200){\circle{.5}}
        \put(1.400,.200){\circle{.5}}
        \put(1.900,.200){\circle{.5}}
        \put(2.400,.200){\circle{.5}}
        \put(2.900,.200){\circle{.5}}
        \put(.650,.633){\circle{.5}}
        \put(1.150,.633){\circle{.5}}
        \put(1.650,.633){\circle{.5}}
        \put(2.150,.633){\circle{.5}}
        \put(2.650,.633){\circle{.5}}
        \put(3.150,.633){\circle{.5}}
        \put(.400,1.066){\circle{.5}}
        \put(.900,1.066){\circle{.5}}
        \put(1.400,1.066){\circle{.5}}
        \put(1.900,1.066){\circle{.5}}
        \put(1.650,1.499){\circle*{.5}}

        \put(5.400,.200){\circle{.5}}
        \put(5.900,.200){\circle{.5}}
        \put(6.400,.200){\circle{.5}}
        \put(6.900,.200){\circle{.5}}
        \put(7.400,.200){\circle{.5}}
        \put(7.900,.200){\circle{.5}}
        \put(5.650,.633){\circle{.5}}
        \put(6.150,.633){\circle{.5}}
        \put(6.650,.633){\circle{.5}}
        \put(7.150,.633){\circle{.5}}
        \put(7.650,.633){\circle{.5}}
        \put(8.150,.633){\circle{.5}}
        \put(5.400,1.066){\circle{.5}}
        \put(5.900,1.066){\circle{.5}}
        \put(6.400,1.066){\circle{.5}}
        \put(7.150,1.133){\circle{.5}}
        \put(6.770,1.440){\circle*{.55}}
        \put(6.770,1.440){\vector(0,-1){.55}}
        \put(7.150,1.133){\vector(1,0){.55}}

        \put(10.400,.200){\circle{.5}}
        \put(10.900,.200){\circle{.5}}
        \put(11.400,.200){\circle{.5}}
        \put(11.900,.200){\circle{.5}}
        \put(12.400,.200){\circle{.5}}
        \put(12.900,.200){\circle{.5}}
        \put(10.650,.633){\circle{.5}}
        \put(11.150,.633){\circle{.5}}
        \put(11.650,.633){\circle{.5}}
        \put(12.150,.633){\circle{.5}}
        \put(12.650,.633){\circle{.5}}
        \put(13.150,.633){\circle{.5}}
        \put(10.400,1.066){\circle{.5}}
        \put(10.900,1.066){\circle{.5}}
        \put(11.400,1.066){\circle{.5}}
        \put(11.900,1.066){\circle*{.5}}
        \put(12.400,1.086){\circle{.5}}

        \put(-.2,4.866){$(a)$}
        \put(-.2,1.866){$(b)$}
      \end{picture}
    \end{picture}
  \end{center}
\caption{
The motion of an atom from the upper terrace to the lower terrace  down 
by $(a)$ jump and $(b)$ exchange mechanism.}
\label{barrier2}
\end{figure}
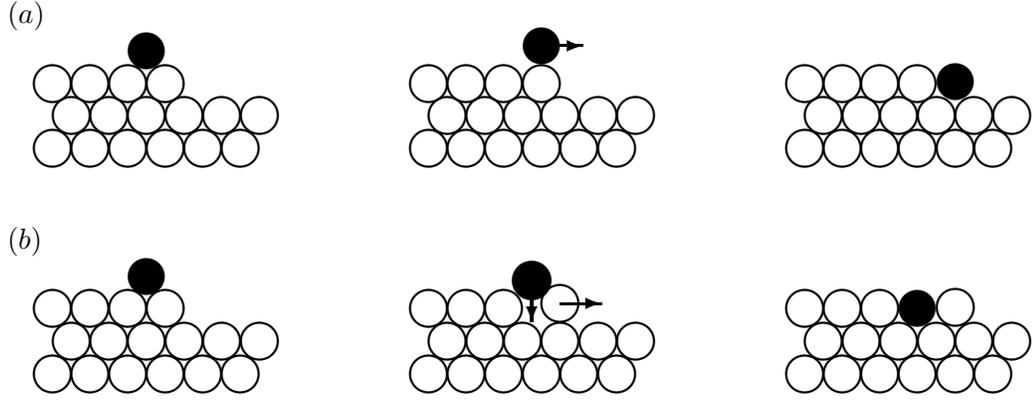 
The exchange process may also be energetically favored for the
diffusion of an adatom from the upper to the lower terrace. As already
mentioned, this diffusion through a normal jump may be hindered by an
additional energy barrier at the step edge. Its origin can be understood in
terms of a bond counting argument. From Fig.~\ref{barrier2}$(a)$ it is
clear that the atom moving to the right toward the step edge (black atom in 
the figure) breaks the bond with the terrace atom on its left before jumping
down the step, but there is no terrace atom on its right to help the diffusion process
in that direction. The absence of this bond generates a further increase in the
energy barrier compared to this of an atom that wanders to the left away from
the step. 
However, for the exchange mechanism the diffusion path is not
affected by such a reduction of the bond number [cf.
Fig.~\ref{barrier2}$(b)$], and the activation barrier for the process may be
lower than for the normal jump.  For some metallic systems [for example, Al on
Al\,(111) and Ag on Ag\,(100)] calculations~\cite{stumpf94,yu96} have shown
that the exchange mechanism is the favored situation for a step-down motion.

\section{{\it Ab initio} Kinetic Monte Carlo}
\label{kmc} 
A severe problem in describing crystal growth is the time scale that
it is too long for MD simulations. MD runs can provide important insight into
elementary microscopic mechanisms but typically they can cover at most 
picoseconds, possibly some nanoseconds. Since growth patterns usually
develop on a time scale of seconds, the inadequacy of MD is evident. Moreover,
the growth structures involve large numbers of particles (from $\sim 10^{2}$
to $\sim 10^{4}$) far beyond the reach of MD (simulations with $\sim 10^2$
atoms are hardly feasible, and only for very short times).  Instead, the
method of choice for studying the spatial and temporal development of growth
is KMC (for more details see Ref.~\cite{ratsch97}, and references therein).
The key idea behind KMC is to describe stochastic processes (such as
deposition, diffusion, desorption, etc.) on the microscopic scales by rates
usually given by:
\begin{equation}
\Gamma^{(j)} = \Gamma_0^{(j)} \exp(-E_{\rm d}^{(j)}/k_{\rm B}T)\quad .
\label{rateexp}
\end{equation} 
Here, $\Gamma_0^{(j)}$ is an effective attempt frequency, $T$
the temperature, $k_{\rm B}$ the Boltzmann constant, and $E_{\rm d}^{(j)}$ is
the energy barrier that needs to be overcome for the event $j$ to take place.
In this way one avoids the explicit calculation of unsuccessful attempts, and
KMC simulations therefore can describe phenomena with time scales of seconds.
Furthermore, large systems can be treated without great difficulties.  The
basic procedure of a KMC simulation can be sketched as follows:
\begin{enumerate}
\item[$1)$] Determination of all processes $j$ that possibly could take place
  with the actual configuration of the system.
\item[$2)$] Calculation of the total rate $R = \sum_j \Gamma^{(j)}$, where the
  sum runs over the possible processes [see step 1)]. Deposition is accounted
  for in this description by the deposition rate $F\quad .$
\item[$3)$] Choose two random numbers $\rho_1$ and $\rho_2$ in the range $(0,1]$.
\item[$4)$] Find the integer number $l$ for which
\begin{equation}
\sum_{j = 1}^{l - 1} \Gamma^{(j)} \leq \rho_1 \, R < \sum_{j = 1}^{l}
\Gamma^{(j)}\quad .
\label{mcalgo}
\end{equation}
\item[$5)$] Let process $l$ to take place.
\item[$6)$] Update the simulation time $t := t + \Delta t$ with $\Delta t=
  -ln(\rho_2)/R\quad .$
\item[$7)$] Go back to step $1)$.
\end{enumerate} 

KMC simulations have been used to study crystal growth of semiconductors
(e.g.~\cite{semimc}) and metals (e.g.~\cite{metmc}).  However, most of these
studies have been based on restrictive approximations.  For example, the input
parameters, such as activation barriers, have been treated as effective
parameters determined rather indirectly, e.g. from the fitting of experimental
quantities. Thus, the connection between these parameters and the microscopic
nature of the processes may be somewhat uncertain.  Often even the surface
structure was treated incorrectly, i.e. the simulation was done on a simple
quadratic lattice while the system of interest had an fcc or bcc structure.
Despite these approximations such studies have provided significant
qualitative quantitative insight into growth phenomena.
The next better approach is to use semi-empirical calculations such as the
embedded atom method or effective medium theory to evaluate the input
parameters for KMC simulations of growth~\cite{eammc}.  The best,
but also most elaborate procedure to obtain these input parameters 
is to exploit the accuracy of DFT, and this is our {\it approach}.

A few words about the important atomistic processes. In view of our
results for Al on Al\,(111), reviewed in Section~\ref{risultati}, we consider
the deposition of monoatomic species with sticking coefficient equal to one,
and the process is quantitatively characterized by the deposition rate $F$.

Concerning adatom diffusion Eq.~(\ref{rateexp}) reflects the idea that
an adatom experiences many stable and metastable sites at the surface,
and that the diffusive motion brings it from one minimum to an
adjacent one on the free energy surface in the configuration space
spanned by the substrate and adatom coordinates.  
Eq.~(\ref{rateexp}) shows that the two ingredients we need are the effective
attempt frequency $\Gamma_0^{(j)}$ and the activation energy barrier
$E_{\rm d}^{(j)}$. Both can be obtained by DFT calculations. The key
quantity for the evaluation of $E_{\rm d}^{(j)}$ is the 
potential energy surface (PES) which is the potential energy surface
seen by the diffusing adatom. The PES is defined as:
\begin{equation}
E^{\rm PES}(X_{\rm ad}, Y_{\rm ad}) = \min_{Z_{\rm ad},\{{\bf R}_I\}}
E^{\rm tot} (X_{\rm ad}, Y_{\rm ad}, Z_{\rm ad},\{{\bf R}_I\}) \quad,
\label{defpes}
\end{equation} 
where $E^{\rm tot} (X_{\rm ad}, Y_{\rm ad}, Z_{\rm ad},\{{\bf
  R}_I\})$ is the ground-state energy of the many-electron system (also
referred as the total energy) at the atomic configuration $(X_{\rm ad}, Y_{\rm
  ad}, Z_{\rm ad},\{{\bf R}_I\})$.  Here, $(X_{\rm ad}, Y_{\rm ad}, Z_{\rm
  ad})$ are the coordinates of the adatom and $\{{\bf R}_I\}$ represent the
set of positions of the substrate atoms. According to Eq.~(\ref{defpes}) the
PES is the minimum of the total energy with respect to the $z$-coordinate of
the adatom $Z_{\rm ad}$ and all coordinates of the substrate atoms $\{{\bf
  R}_I\}$.  Under the assumption of negligible vibrational effects, the minima
of the PES represent stable and metastable sites of the adatom. Moreover, we
consider the dynamics of the electrons and of the nuclei decoupled 
(Born-Oppenheimer approximation)
that is usually well justified for not too high temperatures. We note
that the results of a KMC study of growth will be the same as that of a MD
simulation, provided that the underlying PES is the same; however, as pointed 
out above, KMC can handle significantly longer times.

Now consider all possible paths $l$ to get from one stable or metastable
adsorption site, ${\bf R}_{\rm ad}$, to an adjacent one, ${\bf R}_{\rm ad}{\bf
  '}$.  The energy difference $E_{{\rm d}l}$ between the energy at the saddle
point along $l$ and the energy of the initial geometry is the barrier for this
particular path. If the vibrational energy is small compared to $E_{{\rm
    d}l}$, the diffusion barrier then is the minimum value of all $E_{{\rm
    d}l}$ of the possible paths that connect ${\bf R}_{\rm ad}$ and ${\bf
  R}_{\rm ad}{\bf '}$, and the lowest energy saddle point is called the {\it
  transition state}. Although often only the path with the most favorable
energy barrier is important, it may happen that several paths exist with
comparable barriers or that the PES consists of more than one sheet (e.g.
Ref.~\cite{kley97}).  Then the {\it effective} barrier measured in an
experiment or a molecular dynamics (MD) simulation represents a thermodynamic
average over all possible pathways.

For calculating the second basic quantity, $\Gamma_0^{(j)}$, Transition State
Theory (TST)~\cite{tst} provides a good framework. Within the harmonic
approximation $\Gamma_0^{(j)}$ is proportional to the ratio between the
products of the normal mode frequencies of the system with the adatom at the
equilibrium site and at the saddle point.  The normal mode frequencies are
obtained by diagonalization of the force constant matrix for the system with
the adatom at the equilibrium site and at the transition point. The force
constant matrix reflects the interactions of the adatom with the substrate and
can be derived from DFT calculations~\cite{kley97}.  TST is only valid when
$E_{\rm d}^{(j)}$ is larger than $k_{\rm B}T$. When this requirement is not
fulfilled, the appropriate tool is Molecular Dynamics (MD).  In several works
the attempt frequency is assumed equal for all the processes with a value
that lies in the range of the highest phonon vibration or the Debye frequency.
However, this assumption may not be correct. First, processes with larger
activation barriers may be characterized by a larger attempt frequency than
processes with smaller energy barriers~\cite{boi95}. Moreover, processes that
involve a different number of particles and different bonding configurations
may also be characterized by different attempt frequencies. For examples,
experimental analyses have given $\Gamma_0^{(j)}$ for exchange diffusion larger
by up to two orders of magnitude than for hopping~\cite{exclit}.

An important step in the description of growth is {\it nucleation}. This
is the process by which two or more mobile adatoms meet and
form a stable cluster. A cluster that is formed can suffer two main fates,
provided that other mobile species are available. Either the cluster dissolves
again into smaller constituents or it survives and ultimately grows, as more
adatoms add to its periphery. In the second case the cluster acts as
the initial nucleus for the growth of islands. If agglomerates of $i^{*} + 1$
and more adatoms are stable against break-up, $i^{*}$ is called the size
of the ``critical nucleus''~\cite{PRL_Ratsch}. 

\section{Results: Al/Al\,(111)}\label{risultati}
Having discussed the basic concepts (see Fig.~\ref{processes}) we
describe the results obtained by combining the DFT results for Al on
Al\,(111) obtained by Stumpf and Scheffler~\cite{stumpf94} with a realistic KMC
simulation that takes into account the correct structure of the system.
As already mentioned, the \,(111) surface of an fcc crystal is
characterized by the presence of the two types of close-packed steps,
shown in Fig.~\ref{ste111}.  Experimentally it has been shown that
for Pt\,(111)~\cite{michely93} and Ir\,(111)~\cite{wan91} these two
steps behave quite differently with respect to surface diffusion and
growth. For Al\,(111) the DFT calculations~\cite{stumpf94} predict
that the formation energies of the two steps are different with a
lower energy cost for the formation of the $\{111\}$ faceted step than
of the $\{100\}$ faceted step: 0.232 eV per atom vs. 0.248 eV per
atom. Thus, the two steps have not the same equilibrium
properties. What happens during growth far from equilibrium?

Stumpf and Scheffler~\cite{stumpf94} analyzed microscopic diffusion processes
and in particular determined the activation energies $E_{\rm d}$ for:
\begin{enumerate}
\item[$(i)$] diffusion of a single adatom on the flat surface: $E_{\rm d}$ =
  0.04 eV;
\item[$(ii)$] diffusion from upper to lower terraces which was found to
  proceed by an exchange with a step-edge atom: $E_{\rm d}$ = 0.06 eV for the
  \{100\} step and $E_{\rm d}$ = 0.08 eV for the \{111\} step;
\item[$(iii)$] diffusion parallel to the \{100\} step via hopping: $E_{\rm d}$
  = 0.32 eV (0.44 eV for exchange);
\item[$(iv)$] diffusion parallel to the \{111\} step via exchange: $E_{\rm d}$
  = 0.42 eV (0.48 eV for hopping).
\end{enumerate} 
The DFT calculations give that the binding energy of two
adatoms in a dimer is 0.58 eV~\cite{stumpf94}, and we therefore assume that
dimers, once they are formed, are stable, i.e., they will not dissociate and
$i^* = 1$.  Moreover, we assume that dimers are immobile.  We note that the
reported value for the self-diffusion energy barrier is rather low (0.04
eV)~\cite{stumpf94} and comparable to the energy of optical phonons of
Al\,(111) ($\approx 0.03$ eV). Since these phonon levels are highly populated
at room temperature, simulations for $T > $ 300 K may not be reliable because
the concept of uncorrelated jumps between nearest neighbor sites is no more
valid: In the reached site the adatom cannot cancel the correlation with the
previous jumps since it stays there too shortly in order to be
thermally equilibrated through the interaction with the substrate. We
therefore limited our study to substrate temperatures $T \leq 250$ K.

We adopt periodic boundary conditions, and our rectangular simulation area is
compatible with the geometry of an fcc\,(111) surface. The linear dimensions of the
simulation area range form 1700 \AA\, to 3000~\AA. These dimensions are a
critical parameter, and it is important to ensure that the simulation area is
large enough that artificial correlations of neighboring cells do not affect
the formation of growth patterns.  The mean free path $\lambda$ of a diffusing
adatom before it meets another adatom with possible formation of a nucleation
center or is captured by existing islands is proportional to
$(D/F)$~\cite{stoy} where $D$ and $F$ are the diffusion constant on the flat
surface and the deposition rate, respectively, and should be smaller than the
linear dimension of the simulation array. Because of the fast diffusion of a
single adatom Al on the flat (111) surface our cell is large enough (for the
imposed deposition rate F = 0.08 ML/s) for $T < 150$ K, whereas for $T \geq
150$ K the dimensions of the cell are too small, and the island density is
determined by the simulation array rather than the physics. Nevertheless, the
island shape is determined by local processes (edge diffusions) and is still
meaningful.

In the KMC program two additional insights extracted from the DFT calculations
are included: $(i)$ the attractive interaction between steps and single
adatoms, and $(ii)$ the fact that diffusion processes take place via different
mechanisms (hopping or exchange). Particularly the second point plays an
important role in our investigation, as already mentioned in the
Section~\ref{kmc}: Processes as hopping and exchange may be characterized by
different attempt frequencies~\cite{liu91}. 

\begin{figure}[b]
  \leavevmode
  \includegraphics{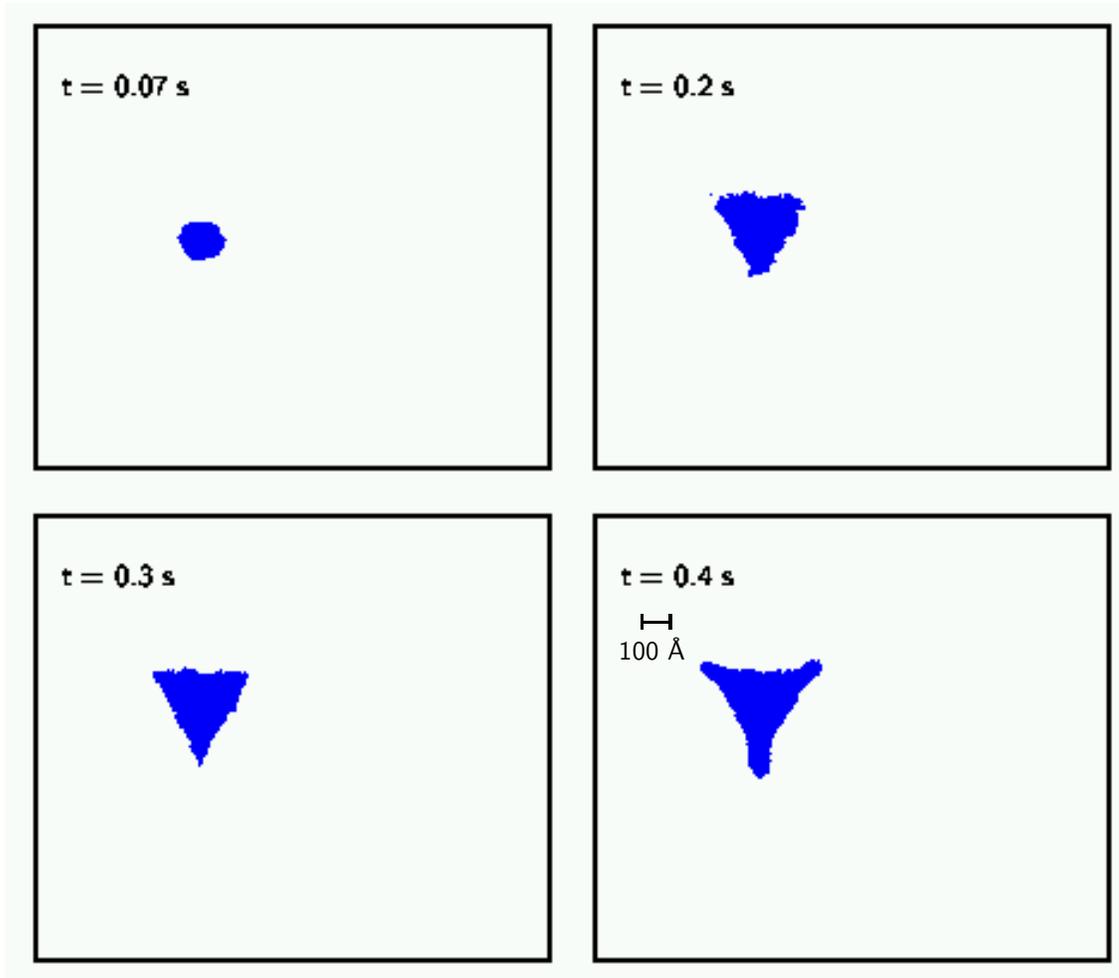}
\unitlength1cm
\begin{picture}(0,0)
\thicklines
\put(9,-8){\line(1,0){0.38}}
\put(9,-8.1){\line(0,1){0.2}}
\put(9.38,-8.1){\line(0,1){0.2}}
\put(8.7,-8.5){\small{\textsf{100 \AA}}}
\end{picture}
\vspace*{13.0cm}
\caption{
A surface area of (1718 $\times$ 1488)~\AA$^2$~ (half of the
  simulation area) at four different substrate temperatures.
The deposition rate was 0.08 ML/s and the coverage in each picture 
is $\Theta$ = 0.04 ML.}
\label{fig.2}
\end{figure}

Results of the {\it ab initio} KMC simulations are collected in
Fig.~\ref{fig.2} for a coverage of $\Theta = 0.04$ ML and $T$ = 50, 150, 200,
and 250 K. With the substrate at $T$ = 50 K during growth the shape of the
islands is highly irregular.  Adatoms which reach a step cannot break away
from it and they even cannot diffuse along the steps.  Thus, at this
temperature ramification occurs into random directions, and island formation
can be understood in terms of the so-called {\it hit and stick} model (see
also Ref.~\cite{san83}).  From Fig.~\ref{fig.2} it appears that at a growth
temperature $T$ = 150~K the islands assumes a triangular shape with their
sides being \{100\} steps.  Increasing the temperature to $T$ = 200~K a
transition from triangular to hexagonal form occurs and for $T$ = 250~K the
islands become triangular again.  However, at this temperature they are mainly
bounded by \{111\} steps.  Another interesting aspect of the simulations is
the appearance of a rough ordered arrangement of the islands obtained at low
temperatures ($T$ = 50 K and 150 K). This arrangement is a kinetic effect and
reflects the fact that the average distance between neighboring islands is
proportional to $(D/F)^{\chi}$ where $\chi$ depends on the size of the
critical nucleus $i^*$.

To understand the island shapes in the temperature regime between 150 and
250~K we consider the mobility of the adatoms along the steps (at such
temperatures the adatoms at the step edges cannot leave the steps): The lower
the migration probability along a given step edge, the higher is the step
roughness and the faster is the growth normal to this step edge.  Thus, this
step edge shortens as a result of the growth kinetics and eventually it may
even disappear.  Since diffusion along the densely packed steps on the (111)
surface (the \{100\} and \{111\} facets) is faster than along steps with any
other orientation this criterion explains the presence of islands which are
mainly bounded by \{100\} or \{111\} steps. The same argument can be extended
to the diffusion along the two close-packed steps and applied to the
triangular islands at $T$ = 150~K, where the energy barrier for the diffusion
along the \{111\} facet is larger and thus the \{100\} steps survive so that
triangular islands with \{100\} sides are obtained. By considering the energy
barriers we would expect only these islands, until the temperature regime for
the thermal equilibrium is reached.
\begin{figure}[t]
\unitlength1cm
\begin{center}
   \begin{picture}(7,6)
      \includegraphics{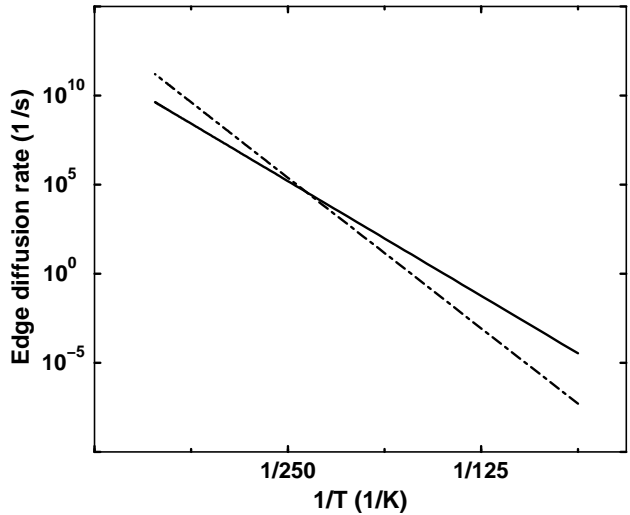}
   \end{picture}
\end{center}
\caption{
  Temperature dependence of the edge diffusion rates (in s$^{-1}$) for atom
  diffusion along the \{100\} step by hopping with $\Gamma_0$ = $2.5 \times
  10^{12}$ s$^{-1}$ (thick solid line), and along the \{111\} step by exchange
  with $\Gamma_0$ = $2.5 \times 10^{14}$ s$^{-1}$ (thin solid line).}
\label{fig4}
\end{figure} 
However, the diffusion of adatoms is not only governed by the
energy barrier but also by the attempt frequency 
$\Gamma_0$ [cf. Eq. (\ref{rateexp})].  For Al/Al\,(111) the effective attempt
frequencies have not been calculated, but the analysis of Ref.~\cite{stumpf94}
proposes that the exchange process should have a larger attempt frequency than
the hopping process.  The results displayed in Fig. ~\ref{fig.2} are obtained
with $\Gamma_0 = 1.0 \times 10^{12}$ s$^{-1}$ for the diffusion on the flat
surface, $\Gamma_0 = 2.5 \times 10^{12}$ s$^{-1}$ for the jump along the
\{100\} step, and $\Gamma_0 = 2.5 \times 10^{14}$ s$^{-1}$ for the exchange
along the \{111\} step.  These effective attempt frequencies are the only
input of the KMC not calculated explicitly by DFT, but were estimated from the
theoretical PES as well as from experimental data for other systems.  In
Fig.~\ref{fig4} the edge diffusion rates along the two steps are plotted as a
function of the reciprocal temperature.  At lower temperatures the energy
barrier dominates the diffusion rate but at $T$ = 250 K the attempt
frequencies start to play a role and lead to faster diffusion along the
\{111\} facet than along the \{100\} one.  Thus, the latter steps disappear
and only triangles with \{111\} sides are present. The roughly hexagonally
shaped islands at $T$ = 200 K are a consequence of the equal
advancement speed for the two steps at that temperature.  
%Obviously, the temperature dependence
%of the growth shapes found in Fig. ~\ref{fig.2} is crucially determined by the
%ratio of the two diffusivities and in particular by the temperature at which
%the two lines of Fig.~ \ref{fig4} cross.  If the difference were only one
%order of magnitude, the crossing would be at a temperature that is too high
%(namely at $T = 505$ K).  The formations of fractals (Fig. ~\ref{fig.2}, upper
%left) and of \{100\} step triangles would still occur.  
Obviously, the importance of the attempt frequencies should receive a better
assessment through accurate calculations, and work in this direction is in
progress.  While no experimental data for Al/Al\,(111) are presently available
we note that a similar sequence of islands as obtained above has been
observed for Pt on Pt\,(111) by Michely {\it et al.}~\cite{michely93}.

\begin{figure}[tb]
  \leavevmode 
\includegraphics{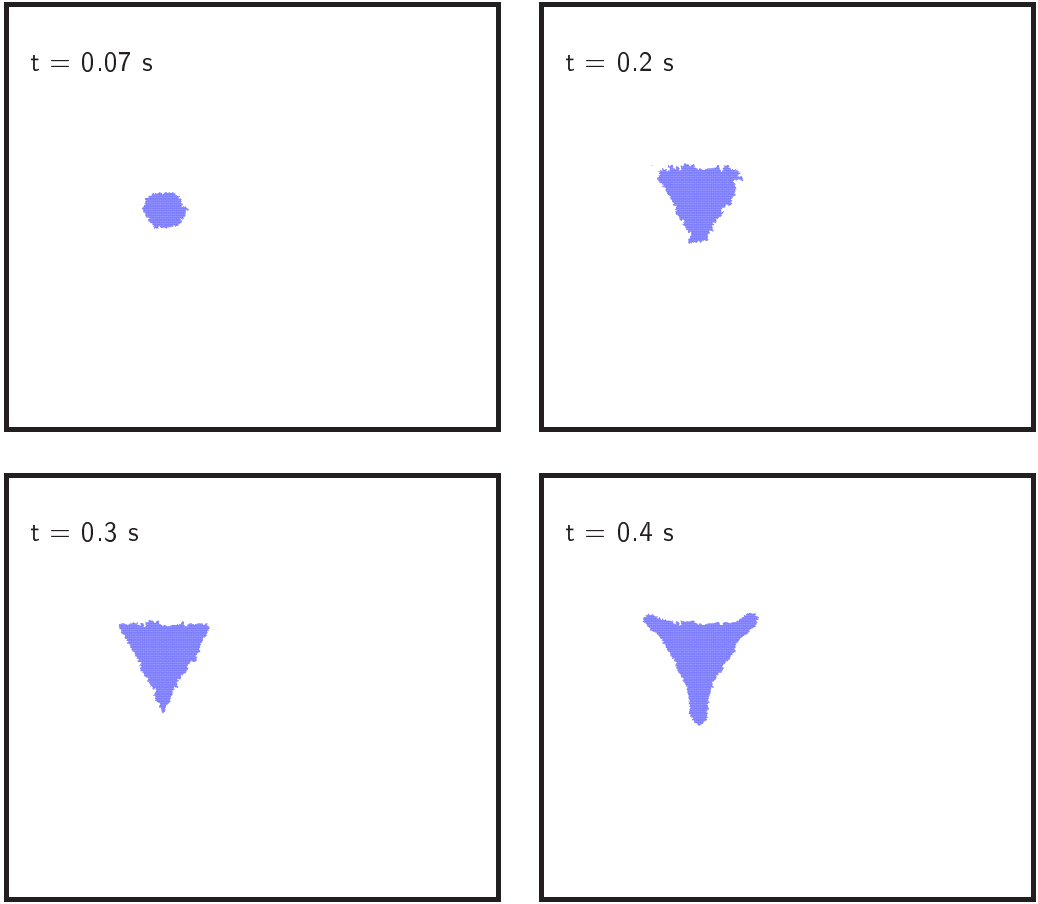} 
\vspace*{13cm}
\caption{
Shape of the islands at $T$ = 250 K as they develop with time (or
coverage). The snapshots refer to
  $t = 0.07$ s, $t = 0.2$ s, $t = 0.3$ s, and $t = 0.4$ s.
The section of the simulation cell that is shown is 1718 $\times$ 1488~\AA$^2$
and the deposition rate is 0.08 ML/s.}
\label{evolu}
\end{figure} 

An interesting point is the evolution of the island shapes with time. This
aspect is related also with a peculiarity of the triangular islands at $T$ =
250 K (cf. Fig.~\ref{fig.2}): They exhibit concave sides. In order to
understand this behavior we have checked the island shape at different times
for the deposition at $T$ = 250 K.  The results are collected in
Fig.~\ref{evolu}.  After $0.07$ s the islands are roughly
hexagonal and upon successive deposition they evolve into a nearly triangular
shape. The longer sides are formed by straight \{111\}-faceted step edges but
short \{100\}-faceted edges can still be identified, at least for $t \leq 0.1$
s.  The latter edges become rougher and progressively disappear. Note that the
development of a triangular island requires $t \sim 0.2$ s. This time is
remarkably long, provided that diffusion of adatoms on the flat surface is
fast at $T$ = 250 K, and it confirms the need of a KMC approach for
investigating in a reliable way the evolution of growth patterns. The shape of
the island is the result of several processes with different time scales that
are accounted for in the KMC scheme. A further confirmation of this collective
action is given by the value of the temperature at which the advancement
speeds for the two steps are practically equal. This corresponds at the
crossing point of the two diffusion rates in Fig.~\ref{fig4} at $T$ = 250 K.
From simulations this temperature is lower ($T$ = 200 K). Going back to the
time evolution, at $t = 0.3$ s the
sides are still nearly straight, but at $t = 0.4$ s the concavities appear.
The corners of the triangles seem to increase their rate of advancement during
deposition. The effect can be understood on the basis of competition between
adatom supply from the flat surface and mass transport along the sides. The
adatom concentration field around an island exhibits the steepest gradient
close to the corners, and the corners of the islands receive an increased flux
of adatoms. When the sides of the islands are not too long, this additional
supply of adatoms is compensated by the mass transport along the steps, i.e.,
the adatoms have a high probability to leave the region around the corners
before the arrival of the successive adatom. For $t = 0.3$ s this scenario
still seems to be true, while at $t = 0.4$ s the island edges are longer and
the mass transport along the sides is not able to compensate the additional
supply of particles at the corners. This means that the probability for a
particle to leave the corner region and to move along the island edge before
being reached by another particle decreases considerably, and the corners
start to grow faster than the sides of the triangles so that the concave shape
develops.

In summary, we have shown how detailed microscopic information can 
be combined with KMC simulations in order to gain insight in
the {\it time} evolution of the structures assembled during growth. As we have
shown for Al on Al\,(111), islands modify their shapes when the substrate
temperature is changed, and the origin of these transitions lies in the
microscopic nature of the diffusion processes occurring onto the
surface. Moreover, growth patterns may develop on a time scale of seconds
that cannot be reached with MD simulations, and the {\it ab initio} KMC
approach seems to be a very powerful tool to overcome size and time limitations 
without loosing the microscopic description or accuracy.

\section{Acknowledgments}
We thank Christian Ratsch (Berlin) for many valuable discussions. One 
of the authors (P.R.) thanks D.E. Wolf (Duisburg) for the kind hospitality and
the useful discussions during his stay at the HLRZ in J\"{u}lich.

\end{document}